\documentstyle[aps,preprint,floats,psfig]{revtex}

\begin{document}
%\psdraft   % only draws the frames for PS figures. faster.
\draft
\preprint{\vbox{\it Submitted to Phys. Lett. B \hfill\rm CU-NPL-1149}}

\title{Magnetic Moments of $\Delta^{++}$ and $\Omega^-$ from QCD Sum Rules}
\author{Frank X. Lee}
\address{Nuclear Physics Laboratory, Department of Physics,
University of Colorado, \\Boulder, CO 80309-0446}
\date{\today}
\maketitle

\begin{abstract}
QCD sum rules for the magnetic moments of 
$\Delta^{++}$ and $\Omega^-$ are derived using the external field method.
They are analyzed by a Monte-Carlo based procedure, using realistic 
estimates of the QCD input parameters.
The results are consistent with the measured values, despite  
relatively large errors that can be attributed mostly 
to the poorly-known vacuum susceptibility $\chi$.
It is shown that a 30\% level accuracy can be achieved 
in the derived sum rules, 
provided the QCD input parameters are improved to the 10\% level.
 
\end{abstract}
\vspace{1cm}
 % Monte Carlo and statistical methods 
\parskip=2mm

The QCD sum rule method~\cite{SVZ79} is a powerful tool in
revealing the deep connection between hadron phenomenology and 
QCD vacuum structure via a few condensate parameters.
The method has been successfully applied to a variety
of problems to gain a field-theoretical understanding 
into the structure of hadrons.
Calculations of the nucleon magnetic moments 
in the approach were first carried out 
in Refs.~\cite{Ioffe84} and~\cite{Balitsky83}.
They were later refined and extended to the entire baryon octet in 
Refs.~\cite{Chiu86,Pasupathy86,Wilson87,Chiu87}.
On the other hand, the magnetic moments of decuplet baryons 
were less well studied within the same approach.
There were previous, unpublished reports in Ref.~\cite{Bely84} 
on $\Delta^{++}$ and $\Omega^-$ magnetic moments.
The magnetic form factor of $\Delta^{++}$
in the low $Q^2$ region was calculated based on a rather different 
technique~\cite{Bely93}.
In recent years, the magnetic moment of $\Omega^-$ has been measured
with remarkable accuracy~\cite{Wallace95}:
$\mu_{\Omega^-}=-2.02\pm0.05\;\mu_N$. 
The magnetic moment of $\Delta^{++}$ has also been extracted from 
pion bremsstrahlung~\cite{Bosshard91}:
$\mu_{\Delta^{++}}=4.5\pm 1.0\;\mu_N$.
The experimental information provides new incentives for 
theoretical scrutiny of these observables.

In this letter, we present an independent calculation of the 
magnetic moments of $\Delta^{++}$ and $\Omega^-$ 
in the QCD sum rule approach.
The goal is two-fold. 
First, we want to find out the applicability 
of the method as an alternative way of understanding 
the measured $\mu_{\Delta^{++}}$ and $\mu_{\Omega^-}$,
and hope to gain some insights into the internal structure
of these baryons from a nonperturbative-QCD perspective.
Second, we want to achieve some realistic understanding of the 
uncertainties involved in such a determination by 
employing a Monte-Carlo based analysis procedure. 
This will help find possible ways for improvement. 
The calculation is algebraically more involved than the octet 
case since one has to deal with the more complex spin structure of 
spin-3/2 particles, but presents no conceptual difficulties.

Consider the two-point correlation function in the QCD vacuum
in the presence of a {\em constant} background 
electromagnetic field $F_{\mu\nu}$: 
\begin{equation}
\Pi_{\alpha\beta}(p)=i\int d^4x\; e^{ip\cdot x}
\langle 0\,|\, T\{\;\eta_{\alpha}(x)\,
\bar{\eta}_{\beta}(0)\;\}\,|\,0\rangle_F,
\label{cf2pt}
\end{equation}
where $\eta_{\alpha}$ is the interpolating field for the propagating baryon.
The subscript $F$ means that the correlation function 
is to be evaluated with an electromagnetic interaction term:
${\cal L}_I = - A_\mu J^\mu$,
added to the QCD Lagrangian.
Here $A_\mu$ is the external electromagnetic potential and 
is given by $A_\mu(x)=-{1\over 2} F_{\mu\nu}x^\nu$ in the fixed-point gauge,
and $J^\mu=e_q \bar{q} \gamma^\mu q$ the quark electromagnetic current. 
The magnetic moments can be obtained by considering 
the {\em linear response} of the correlation function to the external field.
The action of the external field 
is two-fold: it couples directly to the quarks in the baryon interpolating
fields,  and it also polarizes the QCD vacuum. 
The latter can be described by the introduction of vacuum susceptibilities.

The interpolating field is constructed from quark fields, and 
has the quantum numbers of the baryon in question. 
For $\Delta^{++}$ and $\Omega^-$, they are given by
\begin{equation}
\eta_{\alpha}^{\Delta^{++}}(x)=
\epsilon^{abc}\left(u^{aT}(x)C\gamma_\alpha u^b(x)\right) u^c(x),
\end{equation}
\begin{equation}
\eta_{\alpha}^{\Omega^-}(x)=
\epsilon^{abc}\left(s^{aT}(x)C\gamma_\alpha s^b(x)\right) s^c(x),
\end{equation}
where $C$ is the charge conjugation operator, 
and the superscript $T$ means transpose. 
The ability of a interpolating field to 
annihilate the {\em ground state} baryon into the QCD vacuum 
is described by a phenomenological parameter $\lambda_B$ (called current
coupling or pole residue), defined by the overlap
\begin{equation}
\langle 0\,|\,\eta_{\alpha}\,|\,Bps\rangle
=\lambda_B\,u_{\alpha}(p,s),
\end{equation}
where $u_{\alpha}$ is the Rarita-Schwinger spin-vector.

The QCD sum rules are derived by calculating the correlator
in~(\ref{cf2pt}) using Operator-Product-Expansion (OPE), on the one
hand, and matching it to a phenomenological representation, on the other. 
The calculation is similar in spirit to the octet case (see 
Ref.~\cite{Chiu86}, for example), but with the added 
complexity of spin-3/2 structures. 
A direct evaluation led to numerous tensor structures, 
not all of them independent.
The dependencies can be removed by ordering the gamma matrices 
in a specific order (we chose 
$ \hat{p} \gamma_\alpha \gamma_\mu \gamma_\nu \gamma_\beta $ 
where the hat notation means $\hat{p}\equiv p^\mu\gamma_\mu$).
After a lengthy calculation, a sum rule involving only the 
magnetic moment can be isolated at one of the structures:
$ \gamma_\alpha F_{\mu\nu} \sigma_{\mu\nu} p_\beta $.
%Details of the calculation will be presented elsewhere.
Here we give the final results,
for $\Delta^{++}$:
\begin{eqnarray}
& &
  {9\over 28} e_u  L^{4/27} E_1 M^4
+ {3\over 56} e_u  b L^{4/27} 
- {6\over 7} e_u  \chi a^2 L^{12/27}
- {4\over 7} e_u \kappa_v a^2 L^{28/27} {1\over M^2}
\nonumber \\ & &
- {1\over 14} e_u (4\kappa + \xi) a^2 L^{28/27} {1\over M^2}
+ {1\over 4} e_u  \chi m^2_0 a^2 L^{-2/27} {1\over M^2}
+ {1\over 12} e_u  m^2_0 a^2 L^{14/27} {1\over M^4}
\nonumber \\ & &
= \tilde{\lambda}^2_B
\left( {\mu_{\scriptscriptstyle B}\over M^2} + A \right) e^{-M^2_B/M^2},
\label{delpp_we5}
\end{eqnarray}
and for $\Omega^-$:
\begin{eqnarray}
& &
  {9\over 28} e_s  L^{4/27} E_1 M^4
- {15\over 7} e_s f \phi  m_s \chi a L^{-12/27} E_0 M^2
+ {3\over 56} e_s  b L^{4/27} 
- {18\over 7} e_s f m_s a L^{4/27} 
\nonumber \\ & &
- {9\over 28} e_s f \phi (2\kappa + \xi) m_s a L^{4/27} 
- {6\over 7} e_s f^2 \phi \chi a^2 L^{12/27}
- {4\over 7} e_s f^2 \kappa_v a^2 L^{28/27} {1\over M^2}
\nonumber \\ & &
- {1\over 14} e_s f^2 \phi (4\kappa + \xi) a^2 L^{28/27} {1\over M^2}
+ {1\over 4} e_s f^2 \phi \chi m^2_0 a^2 L^{-2/27} {1\over M^2}
\nonumber \\ & &
- {9\over 28} e_s f m_s m^2_0 a L^{-10/27} {1\over M^2}
+ {1\over 12} e_s f^2 m^2_0 a^2 L^{14/27} {1\over M^4}
\nonumber \\ & &
= \tilde{\lambda}^2_B
\left( {\mu_{\scriptscriptstyle B}\over M^2} + A \right) e^{-M^2_B/M^2}.
\label{omeg_we5}
\end{eqnarray}
In these equations, 
the magnetic moment $\mu_B$ is given in particle's natural magnetons.
The various symbols are defined as follows. 
The condensates are represented by
$a=-(2\pi)^2\,\langle\bar{u}u\rangle$,
$b=\langle g^2_c\, G^2\rangle$, $\langle\bar{u}g_c\sigma\cdot G
u\rangle=-m_0^2\,\langle\bar{u}u\rangle$,
and the coupling $\tilde{\lambda}_B=(2\pi)^2\lambda_B$. 
The factors $e_u=2/3$ and $e_s=-1/3$ are quark charges in
units of electric charge.
The vacuum susceptibilities $\chi$, $\kappa$ and $\xi$ are defined by 
$\langle\bar{q} \sigma_{\mu\nu} q\rangle_F \equiv
e_q \chi \langle\bar{q}q\rangle F_{\mu\nu}$,
$\langle\bar{q} g_c G_{\mu\nu} q\rangle_F \equiv
e_q \kappa \langle\bar{q}q\rangle F_{\mu\nu}$, 
and
$\langle\bar{q} g_c \epsilon_{\mu\nu\rho\lambda} G^{\rho\lambda} \gamma_5 
q\rangle_F \equiv i e_q \xi \langle\bar{q}q\rangle F_{\mu\nu}$.
The parameters $f=\langle\bar{s}s\rangle/\langle\bar{u}u\rangle
=\langle\bar{s}g_c\sigma\cdot G s\rangle
/\langle\bar{u}g_c\sigma\cdot G u\rangle$
and $\phi=\chi_s/\chi=\kappa_s/\kappa=\xi_s/\xi$
account for flavor symmetry breaking of the strange quark.
Possible violation of the four-quark condensate is considered by 
the parameter $\kappa_v$ as defined in 
$\langle\bar{u}u\bar{u}u\rangle=\kappa_v\langle\bar{u}u\rangle^2$.
The anomalous dimension corrections of the various operators 
are taken into account via the factor $L=\left[{\alpha_s(\mu^2)/
\alpha_s(M^2)}\right] =\left[{\ln(M^2/\Lambda_{QCD}^2)/
\ln(\mu^2/\Lambda_{QCD}^2)}\right]$, where $\mu=500$ MeV is the
renormalization scale and $\Lambda_{QCD}$ is the QCD scale parameter.  
As usual, the pure excited state contributions are modeled using terms 
on the OPE side surviving $M^2\rightarrow \infty$ under the assumption
of duality, and are represented by the factors 
$E_n(x)=1-e^{-x}\sum_n{x^n/n!}$ with $x=w^2/M_B^2$
and $w$ an effective continuum threshold. 
The parameter $A$ accounts for all contributions from 
the transitions caused by the external field 
between the ground state and the excited states.
Such contributions are {\em not} exponentially suppressed relative to the 
ground state double pole and must be included.
The presence of such contributions is a general feature of the external
field technique. 

Let us note in passing that 
since $\Omega^-$ and $\Delta^{++}$ are simply related by the
interchange of quark flavors $u\leftrightarrow s$,
one can verify that sum rule~(\ref{omeg_we5}) reduces to 
sum rule~(\ref{delpp_we5}) if one sets $e_s\rightarrow e_u$, 
$m_s=0$, $f=1$, and $\phi=1$. 

To analyze the sum rules, we use a Monte-Carlo based
procedure first developed in Ref.~\cite{Derek96}.
The basic steps are as follows.
First, the uncertainties in the QCD input parameters are assigned.
Then, randomly-selected, Gaussianly-distributed sets are generated,
from which an uncertainty distribution in the OPE can
be constructed.  Next, a $\chi^2$ minimization is applied to the sum
rule by adjusting the phenomenological fit parameters.  This is done
for each QCD parameter set, resulting in distributions for
phenomenological fit parameters, from which errors are derived.
Usually, 100 such configurations are sufficient for getting stable
results. We generally select 1000 sets which help resolve more subtle
correlations among the QCD parameters and the phenomenological fit
parameters.

The Borel window over which the two sides
of a sum rule are matched is determined by the following two criteria.
First, {\em OPE convergence}: the highest-dimension-operators
contribute no more than 10\% to the QCD side.
Second, {\em ground-state dominance}: excited
state contributions should not exceed more than 50\% of the 
phenomenological side.
The first criterion effectively establishes a lower limit, 
the second an upper limit.
Those sum rules which do not have a valid Borel window under these 
criteria are considered unreliable and therefore discarded.  

The QCD input parameters and their uncertainty assignments are 
given as follows~\cite{Lee97a}.
The condensates are taken as $a=0.52\pm0.05$ GeV$^3$, 
$b=1.2\pm0.6$ GeV$^4$, and $m^2_0=0.72\pm0.08$ GeV$^2$.
For the factorization violation parameter,
we use $\kappa_v=2\pm 1$ and $1\leq \kappa_v \leq 4$.
The QCD scale parameter is restricted to $\Lambda_{QCD}$=0.15$\pm$0.04 GeV.
The vacuum susceptibilities have been estimated in 
studies of nucleon magnetic moments~\cite{Ioffe84,Balitsky83,Chiu86},
but the values vary in a wide range depending on the method used.
Here we take their median values with 50\% uncertainties:
$\chi=-6.0\pm 3.0$ GeV$^{-2}$,
$\kappa=0.75\pm 0.38$, and $\xi=-1.5\pm 0.75$.
Note that $\chi$ is almost an order of magnitude 
larger than $\kappa$ and $\xi$, and is the most important of the three.
The strange quark parameters are placed at $m_s=0.15\pm 0.02$ GeV,
$f=0.83\pm0.05$ and $\phi=0.60\pm0.05$~\cite{Pasupathy86}. 
These uncertainties are assigned conservatively and in accord with the
state-of-the-art in the literature. 
While some may argue that some
values are better known, others may find that the errors are
underestimated.  In any event, one will learn how the uncertainties in
the QCD parameters are mapped into uncertainties in the
phenomenological fit parameters.

To illustrate how well a sum rule works, we first cast it into 
the subtracted form, $\Pi_S=\tilde{\lambda}^2_B \mu_B e^{-M^2_B/M^2}$,
then plot the logarithm of the absolute value of the two sides
against the inverse of $M^2$.
In this way, the right-hand side will appear as a straight line
whose slope is $-M_B^2$ and whose intercept with the y-axis gives
some measure of the coupling strength and the magnetic moment.
The linearity (or deviation from it) of the left-hand side gives 
an indication of OPE convergence, and information 
on the continuum model and the transitions.

To extract the magnetic moments, a two-stage fit was performed. 
First, the corresponding chiral-odd mass sum rule, as obtained 
previously in  Ref.~\cite{Lee97a}, was fitted to get the mass
$M_B$, the coupling $\tilde{\lambda}_B^2$ and the continuum threshold
$w_1$.  Then, $M_B$ and $\tilde{\lambda}_B^2$ were used in the 
magnetic moment sum rule for a three-parameter fit:
the transition strength $A$, the continuum threshold $w_2$, 
and the magnetic moment $\mu_B$.
Note that $w_1$ and $w_2$ are not necessarily the same.
We impose a physical constraint on both $w_1$ and $w_2$ 
requiring that they are larger than the mass,
and discard QCD parameter sets that do not satisfy this condition.
The above procedure is repeated for each QCD parameter set until a
certain number of sets are reached.
In the actual analysis of sum rules~(\ref{delpp_we5}) 
and~(\ref{omeg_we5}), however, we found that such a full search 
was unsuccessful: 
the search algorithm consistently returned $w_2$
either zero or smaller than $M_B$. This signals insufficient information 
in the OPE to completely resolve the spectral parameters.
To proceed, we fixed $w_2$ at $w_1$, which is a 
reasonable and commonly adopted choice in the literature, 
and searched for $A$ and $\mu_B$.

\begin{figure}[tbh]
\centerline{\psfig{file=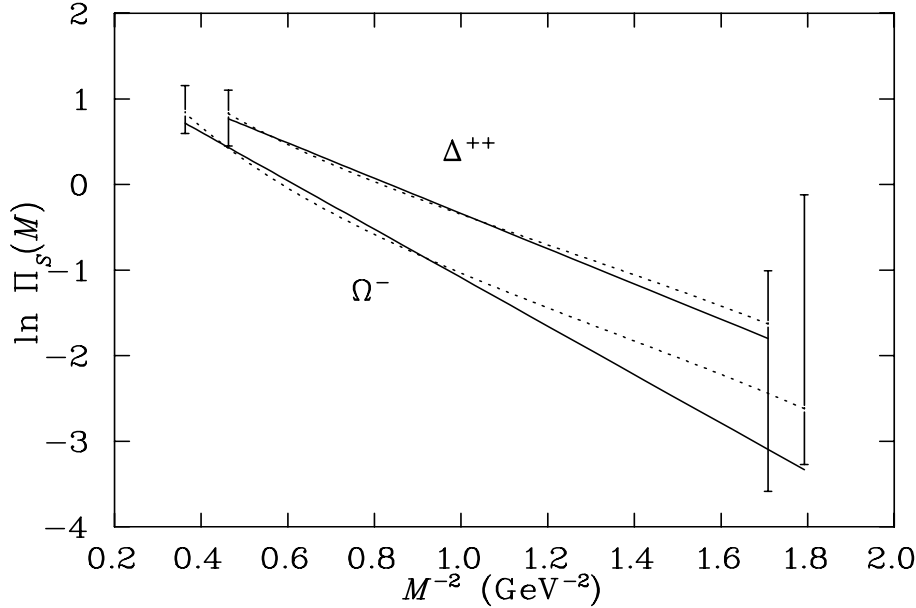,height=8cm,width=12cm,angle=90}}
\vspace{1cm}
\caption{Monte-Carlo fits of the magnetic moment 
sum rules~(\ref{delpp_we5}) and~(\ref{omeg_we5}).
Each sum rule is searched independently.
The solid line corresponds  to the ground state contribution, 
the dotted line the rest of the contributions 
(OPE minus continuum minus transition).
The error bars are only shown at the two ends for clarity.}
\label{rhslhs_do}
\end{figure}

Fig.~\ref{rhslhs_do} shows the match for the sum 
rules~(\ref{delpp_we5}) and~(\ref{omeg_we5}).
The extracted results are given in Table~\ref{tabdo}.
%
%%%%%%%%%%%%%%%%%%%%%%%%%%%%%%%%%%%%%%%%%%%%%%%%%%%%%%%
%%%%%%%%%%%%%%%%%%%%%%%%%%%%%%%%%%%%%%%%%%%%%%%%%%%%%%%
%
%
%%%%%%%%%%%%%%%%%%%%%%%%%%%%%%%%%%%%%%%%%%%%%%%%%%%%%%%%%%%%%%
\begin{table}[tbh]
\caption{Monte-Carlo analysis of the QCD sum rules 
for the magnetic moments of $\Delta^{++}$ and $\Omega^{-}$.
The third column represents the percentage contribution of the 
excited states and transitions 
to the phenomenological side at the lower end of the Borel region 
(it increases to 50\% at the upper end).  
The second row for each sum rule corresponds to results 
with reduced, uniform 10\% errors assigned to all the QCD input parameters.
The uncertainties were obtained from 
consideration of 1000 QCD parameter sets.
In the event that the resultant distribution is not Gaussian, 
the median and asymmetric deviations are reported.}
\label{tabdo}
\begin{tabular}{rccccc}
Sum Rule & Region & Cont      & $w$ & A & $\mu_B$ \\
& (GeV) & (\%) & (GeV) & (GeV$^{-2}$)   &  $(\mu_N)$ 
\\ \hline
(\protect\ref{delpp_we5}) for $\Delta^{++}$
& 0.765 to 1.47 & 9& 1.65  & $0.53\pm 0.77$ & $3.60^{+3.68}_{-3.55}$ \\
& 0.765 to 1.47 & 8& 1.65  & $0.35\pm 0.37$ & $4.13^{+1.40}_{-1.18}$ \\
(\protect\ref{omeg_we5}) for $\Omega^{-}$
&0.747 to 1.66 & 7& 2.30 & $-0.15\pm 0.14$ & $-1.25^{+1.12}_{-1.17}$ \\
&0.747 to 1.66 & 6& 2.30 & $-0.10\pm0.05$  & $-1.49^{+0.40}_{-0.49}$ 
\end{tabular}
\end{table}
%%%%%%%%%%%%%%%%%%%%%%%%%%%%%%%%%%%%%%%%%%%%%%%%%%%%%%%%%%%%%%
%
Relatively large errors in the magnetic moments are found, 
approaching 100\%.
But the sign and order of magnitudes are unambiguous when 
compared to the measured values.
The situation is consistent with a previous finding on $g_A$~\cite{Lee97}
regarding three-point functions.
It is interesting to observe that
the distribution of errors is not uniform throughout
the Borel window,  with the largest errors at the lower end where 
the power corrections are expected to become more important.
The quality of the match deteriorates for $\Omega^{-}$ in this region,
signaling probably insufficient convergence of the OPE.
The match for $\Delta^{++}$ is good in the entire Borel region, 
despite the large uncertainties. 
To gain some idea on how the uncertainties change with the input,
we also analyzed the sum rules by adjusting the error estimates
individually.
We found large sensitivities to the susceptibility $\chi$.
In fact, most of the errors came from the uncertainty in $\chi$.
We also tried with reduced error estimates on all
the QCD input parameters: 10\% relative errors uniformly.
The results are given in Table~\ref{tabdo} as a second entry.
It leads to about 30\% accuracy on the magnetic moments.
Further improvement of the accuracy by 
reducing the input errors is beyond the capability of these sum rules
as it will lead to 
unacceptably large $\chi^2/N_{DF}$~\cite{Derek96}.
For that purpose, one would have to resort to finding sum rules 
that depend less critically on $\chi$ and have better convergence 
properties.

A comparison with those from other calculations and the experimental
data is compiled in Table~\ref{comp}. The results with 10\% errors 
from the QCD sum rule method are used in the comparison.
They are consistent with data within errors, 
although the central value for $\Omega^-$ is somewhat underestimated. 
The result for $\Delta^{++}$ is consistent with other calculations,
while for $\Omega^-$ it is closer to lattice QCD calculations.
%
%%%%%%%%%%%%%%%%%%%%%%%%%%%%%%%%%%%%%%%%%%%%%%%%%%%%%%%
\begin{table}[tb]
\caption{Comparisons of magnetic moments from
various calculations: this work (QCDSR),
lattice QCD (Latt)~\protect\cite{Derek92},
chiral perturbation theory ($\chi$PT)~\protect\cite{Butler94},
light-cone relativistic constituent quark model (RQM)~\protect\cite{Schlumpf93},
simple non-relativistic constituent quark model (NQM),
chiral quark-soliton model ($\chi$QSM)~\protect\cite{Kim97}.
All results are expressed in units of nuclear magnetons.}
\label{comp}
\begin{tabular}{lccccccc}
Baryon & Exp.  & QCDSR & Latt & $\chi$PT &  RQM & NQM & $\chi$QSM
\\ \hline
$\Delta^{++}$
&4.5 $\pm$ 1.0       &  4.13 $\pm$ 1.30 & 4.91 $\pm$ 0.61 
& 4.0 $\pm$ 0.4 & 4.76 & 5.56 & 4.73 \\
$\Omega^{-}$
& -2.024 $\pm$ 0.056 & -1.49 $\pm$ 0.45 & -1.40 $\pm$ 0.10 
& ---                  &-2.48 & -1.84 & -2.27\\
\end{tabular}
\end{table}
%%%%%%%%%%%%%%%%%%%%%%%%%%%%%%%%%%%%%%%%%%%%%%%%%%%%%%%
%

In conclusion, we have demonstrated that the magnetic moments of 
$\Delta^{++}$ and $\Omega^{-}$ can be understood from the QCD 
sum rule approach, despite large errors that can be traced to
the uncertainties the QCD input parameters.
A 30\% accuracy can be achieved in the derived sum rules
with improved estimates of the QCD input parameters, 
preferably on the 10\% accuracy level.
In particular, large sensitivities to the vacuum susceptibility $\chi$ 
are found. Better estimate of this parameter is needed.
Extension of the calculations to other members of the 
decuplet appears straightforward, and is under way~\cite{Lee97c}.
There we hope to address the issues involved in greater detail.

It is a pleasure to thank D.B. Leinweber for providing 
an original version of his Monte-Carlo
analysis program and for helpful discussions.
This work was supported in part by 
U.S. DOE under Grant DE-FG03-93DR-40774.


\begin{references}

\bibitem{SVZ79} M.A. Shifman, A.I. Vainshtein and Z.I. Zakharov,
Nucl. Phys. {\bf B147}, 385, 448 (1979).

\bibitem{Ioffe84} B.L. Ioffe and A.V. Smilga,
Phys. Lett. {\bf B133}, 436 (1983);
Nucl. Phys. {\bf B232}, 109 (1984).

\bibitem{Balitsky83} B.L. Ioffe and A.V. Smilga,
Phys. Lett. {\bf B129}, 328 (1983).

\bibitem{Chiu86} C.B. Chiu, J. Pasupathy, S.L. Wilson,
Phys. Rev. {\bf D33 }, 1961 (1986).

\bibitem{Pasupathy86} J. Pasupathy, J.P. Singh, S.L. Wilson, and C.B.  Chiu,
Phys. Rev. {\bf D36 }, 1442 (1986).

\bibitem{Wilson87} S.L. Wilson, J. Pasupathy, C.B. Chiu,
Phys. Rev. {\bf D36 }, 1451 (1987).

\bibitem{Chiu87} C.B. Chiu, S.L. Wilson, J. Pasupathy, and J.P. Singh,
Phys. Rev. {\bf D36 }, 1553 (1987).

\bibitem{Bely84} V.M. Belyaev, 
preprint ITEP-118 (1984); ITEP report (1992), unpublished.

\bibitem{Bely93} V.M. Belyaev, 
preprint CEBAF-TH-93-02, hep-ph/9301257.

\bibitem{Wallace95} N.B. Wallace {\it et al.},
Phys. Rev. Lett. {\bf 74}, 3732 (1995).

\bibitem{Bosshard91} A. Bosshard  {\it et al.},
Phys. Rev. {\bf D44}, 1962 (1991).

\bibitem{Derek96}D.B. Leinweber,
Ann. of Phys. (N.Y.) {\bf 254}, 328 (1997).

\bibitem{Lee97a} F.X. Lee, 
%"Predicative Ability of QCD Sum Rules for Decuplet baryons", 
preprint CU-NPL-1147, hep-ph/9707332.

\bibitem{Lee97} F.X. Lee, D.B. Leinweber, and X. Jin, 
Phys. Rev. {\bf D55}, 4066 (1997).

\bibitem{Derek92} D.B. Leinweber, T. Draper, and R.M. Woloshyn,
Phys. Rev. {\bf D46}, 3067 (1992).

\bibitem{Butler94} M.N. Butler, M.J. Savage,  and R.P. Springer,
Phys. Rev. {\bf D49}, 3459 (1994).

\bibitem{Schlumpf93} F. Schlumpf,
Phys. Rev. {\bf D48}, 4478 (1993).

\bibitem{Kim97} H.C. Kim, M. Praszalowicz, and K. Goeke,     
hep-ph/9706531.

\bibitem{Lee97c} F.X. Lee, in preparation.

\end{references}
\end{document}